\pdfoutput=1

\documentclass[preprint,numbers,sort&compress,5p]{elsarticle}

\usepackage{graphicx}
\usepackage{amsmath,amssymb}
\usepackage{mathtools}
\usepackage{natbib}
\usepackage{hyperref}
\hypersetup{pdftitle = The title of my PDF, pdfauthor = My name, pdfsubject= The subject, pdfkeywords = keyword1 keyword2 keyword3}
\hypersetup{colorlinks = true, linkcolor = blue, anchorcolor = red, citecolor = blue, filecolor = red, pagecolor = red, urlcolor = blue}

\journal{Physics Letters A}

\begin{document}

\begin{frontmatter}
\title{Dynamical analysis of bounded and unbounded orbits in a generalized H\'enon-Heiles system}

\author[fld]{F. L. Dubeibe\corref{cor1}}
\ead{fdubeibe@unillanos.edu.co}

\author[fld]{A. Ria\~no-Doncel}
\ead{angelica.riano@unillanos.edu.co}

\author[eez]{Euaggelos E. Zotos}
\ead{evzotos@physics.auth.gr}

\cortext[cor1]{Corresponding author}

\address[fld]{Facultad de Ciencias Humanas y de la Educaci\'on, Universidad de los Llanos, Villavicencio, Colombia}

\address[eez]{Department of Physics, School of Science, Aristotle University of Thessaloniki, GR-541 24, Thessaloniki, Greece}

\begin{abstract}
The H\'enon-Heiles potential was first proposed as a simplified version of the gravitational potential experimented by a star in the presence of a galactic center. Currently, this system is considered a paradigm in dynamical systems because despite its simplicity exhibits a very complex dynamical behavior. In the present paper, we perform a series expansion up to the fifth-order of a potential with axial and reflection symmetries, which after some transformations, leads to a generalized H\'enon-Heiles potential. Such new system is analyzed qualitatively in both regimes of bounded and unbounded motion via the Poincar\'e sections method and plotting the exit basins. On the other hand, the quantitative analysis is performed through the Lyapunov exponents and the basin entropy, respectively. We find that in both regimes the chaoticity of the system decreases as long as the test particle energy gets far from the critical energy. Additionally, we may conclude that despite the inclusion of higher order terms in the series expansion, the new system shows wider zones of regularity (islands) than the ones present in the H\'enon-Heiles system.
\end{abstract}


\end{frontmatter}

\section{Introduction}
\label{sec1}

The H\'enon-Heiles Hamiltonian \cite{Henon1964} is considered a representative model  for time-independent Hamiltonian systems with two degrees of freedom and thus an essential topic in many textbooks on Nonlinear dynamics, see for example \cite{Tabor1989,Hilborn2000,Gutzwiller2013}. The main reasons for that are its simple analytical form and at the same time its far from trivial dynamics. Such system was originally formulated to shed lights on the question: does an axisymmetric potential admit a third isolating integral of motion? Nowadays it can be considered one of the most cited works in the field of complex systems, where a huge amount of research has been devoted to discriminate between regular and chaotic motion or to study the escape dynamics of orbits, see e.g. \cite{Aguirre2009,Barrio2009,Conte2005,deMoura1999,Fordy1991,Wojciechowski1984,Zotos2015,Zotos2017}. Although its application was first oriented to the field of galactic dynamics, its applications include semiclassical and quantum mechanics \cite{Jaffe1982,Feit1984,Hamilton1986}.

Some modifications to the original H\'enon-Heiles system (henceforth HH-system) have been proposed by different authors, e.g, adding dissipation terms \cite{Seoane2007},  introducing white noise \cite{Seoane2008,Seoane2010}, or including forcing terms \cite{Blesa2012,Coccollo2013}, just to name a few. On the other hand, to our knowledge, the only formal derivation of a generalized H\'enon-Heiles potential has been carried out by Verhulst \cite{Verhulst1979}, who performed a series expansion up to the fourth-order of a potential with axial and reflection symmetries, finding that his potential differs from the one of H\'enon-Heiles by the presence of the quartic polynomial $a_{1} x^{4} + a_{2} x^{2} y^{2}+ a_{3} y^{4}$. In addition, it was found that as the area enclosed by the zero velocity curve tends to zero an additional isolating integral can generally be derived \cite{Verhulst1979}. The Verhulst potential has been used, for example, to study the orbital structure near the center of a triaxial galaxy with an analytic core \cite{Zeeuw1985}, the correlation between the Lyapunov exponents and the size of the chaotic regions in the surface of section \cite{Cleary1989}, or the escape regions in a quartic potential \cite{Barbanis1990}.

Following the procedure outlined in Refs. \cite{Verhulst1979} and \cite{Contopoulos2002}, in the present paper we perform a series expansion of an axisymmetric potential up to the fifth-order, such that the resulting Hamiltonian has as particular cases the H\'enon-Heiles and Verhulst systems (abusing terminology, all along the paper we will call the new potential generalized H\'enon-Heiles system or simply GHH-system). In order to analyze the gradual transition of the dynamics from the HH-system to the GHH-system, we shall introduce a factor $\delta$ multiplying the quartic and quintic polynomial. Since it is a well-known fact that some types of Hamiltonian systems have a finite energy of escape at which the equipotential surfaces give place to exit channels, here we aim to study the dynamics of the GHH-system not only for energy values below the escape energy (bounded orbits) but also above this threshold value (unbounded orbits).

The paper is organized as follows: In section \ref{sec2} we calculate the fifth-order series expansion of the potential, and then we write down the Hamiltonian of the GHH-system with their respective equations of motion. Next, we calculate the critical values of energy for bounded and unbounded motions as a function of $\delta$. In section \ref{sec3}
we study the dynamics of bounded orbits through the Poincar\'e sections and Lyapunov exponents as a function of the total orbital energy and the parameter $\delta$. Unbounded orbits are analyzed by means of the exit basins and the basin entropy in section \ref{sec4}. Finally, our paper ends with section \ref{sec5}, where the conclusions are presented.

\section{Approximate Potential}
\label{sec2}

The most general Hamiltonian in cylindrical coordinates $(r,\theta,z)$ for a test particle in an axisymmetric potential can be written as
\begin{equation}\label{eq:2.1}
{\cal H}= \frac{1}{2}(\dot{r}^{2}+\dot{z}^{2})+V(r,z)+\frac{L_{z}^{2}}{2 r^{2}},
\end{equation}
with $L_{z}=r^{2} \dot{\theta}$ the component of angular momentum about the $z$-axis.

Let us define the effective potential as $V_{\rm eff}(r,z)=V(r,z)+L_{z}^{2}/2 r^{2}$, which as a minimum at $(r,z)=(r_{0},0)$, where
\begin{equation}\label{eq:2.2}
\frac{\partial V}{\partial r}-\frac{L_{z}^{2}}{r^{3}}=0
\end{equation}
and $r_0$ corresponds to the radius of a circular orbit on the symmetry plane. By imposing reflection symmetry in $z$, the potential $V(r,z)$ must be an even function in $z$ such that all its odd-powered derivatives are odd functions. Expanding the effective potential as a Taylor series around $(r_{0},0)$ up to the 5th-order, we find
\begin{align}\label{eq:2.3}
V_{\rm eff}(r,z) &\approx z^4 (a_{1}+b_{2} \xi )+z^2 \left(a_{2} \xi ^2+b_{3} \xi^3+\frac{\omega_{2}^2}{2}-\xi  \epsilon \right) \nonumber\\
&+ a_{3} \xi^4 - \frac{\beta  \xi ^3}{3}+b_{1} \xi^5+\frac{\xi^2 \omega_{1}^2}{2},
\end{align}
where we have used the fact that an odd function is zero at the origin, we omitted the constant terms, and the remaining constants have been set as follows
\begin{eqnarray}\label{eq:2.4}
&&\xi=r-r_{0}, \nonumber\\
&&
\omega_{1}^{2}=\left. \frac{\partial^{2} V_{\rm eff}}{\partial r^2}\right\vert_{(r_0,0)}, \quad \omega_{2}^{2}=\left. \frac{\partial^{2} V_{\rm eff}}{\partial z^2}\right\vert_{(r_0,0)}, \quad  \nonumber\\
&&
\epsilon= -\left. \frac{1}{2}\frac{\partial^{3} V_{\rm eff}}{\partial r \partial z^2}\right\vert_{(r_0,0)}, \quad
\beta=-\left. \frac{1}{2}\frac{\partial^{3} V_{\rm eff}}{\partial r^3}\right\vert_{(r_0,0)}, \quad \nonumber\\
&&
a_{1}= \left. \frac{1}{24}\frac{\partial^{4} V_{\rm eff}}{\partial z^4}\right\vert_{(r_0,0)}, \quad
a_{2}=\left. \frac{1}{4}\frac{\partial^{4} V_{\rm eff}}{\partial r^2 \partial z^2}\right\vert_{(r_0,0)}, \quad \nonumber\\
&&
a_{3}= \left. \frac{1}{24}\frac{\partial^{4} V_{\rm eff}}{\partial r^4}\right\vert_{(r_0,0)}, \quad
b_{1}=\left. \frac{1}{120}\frac{\partial^{5} V_{\rm eff}}{\partial r^5}\right\vert_{(r_0,0)}, \quad \nonumber\\
&&
b_{2}=\left. \frac{1}{24}\frac{\partial^{5} V_{\rm eff}}{\partial r \partial z^4}\right\vert_{(r_0,0)}, \quad
b_{3}=\left. \frac{1}{12}\frac{\partial^{5} V_{\rm eff}}{\partial r^3 \partial z^2}\right\vert_{(r_0,0)}.
\end{eqnarray}

With the aim to compare the new potential (\ref{eq:2.3}) with the one derived by H\'enon and Heiles \cite{Henon1964}, we replace $z\rightarrow x$, $\xi\rightarrow y$, and choose the values
$a_{1}=a_{3}=b_{1}=-b_{2}=-b_{3}=-\delta$, $a_{2}=-2\delta$, $\omega_{1}=\omega_{2}=\beta=-\epsilon=1$, such that the new Hamiltonian takes the form\footnote{As can be easily noted, setting $\delta=0$ in Eq. (\ref{eq:2.5}) we get the well-known H\'enon-Heiles system \cite{Contopoulos2002}.}
\begin{align}\label{eq:2.5}
{\cal H} &= \frac{1}{2} \left(\dot{x}^2+\dot{y}^2\right)+\frac{1}{2} \left(x^2+y^2\right)+x^2 y-\frac{y^3}{3} \nonumber\\
&+ \delta\left[x^4 y+x^2 y^3-y^5-\left(x^2+y^2\right)^2\right].
\end{align}

The equations of motion derived from Hamilton's equations read as
\begin{eqnarray}
\dot{x}&=&p_{x},\label{eqm1}\\
\dot{y}&=&p_{y},\\
\dot{p_{x}}&=&-\frac{\partial {\cal H}}{\partial x},\\
\dot{p_{y}}&=&-\frac{\partial {\cal H}}{\partial y}.\label{eqm4}
\end{eqnarray}
where $p_{x}$ and $p_{x}$ are the canonical conjugate momenta to $x$ and $y$, respectively. Since the total energy is conserved (${\cal H}=E=$ constant), the orbital motion is restricted to the region
\begin{align}\label{eq:2.6}
E &\geq \frac{1}{2} \left(x^2+y^2\right)+x^2 y-\frac{y^3}{3} \nonumber\\
&+ \delta\left[x^4 y+x^2 y^3-y^5-\left(x^2+y^2\right)^2\right].
\end{align}

Depending on the value of the parameter $\delta$, the dynamical system (\ref{eqm1}-\ref{eqm4}) has a given number of real fixed points at which
\begin{equation}
\frac{\partial U(x,y)}{\partial x}= \frac{\partial U(x,y)}{\partial y}=0,
\end{equation}
where
\begin{align}\label{eq:2.7}
U(x,y) &= \frac{1}{2} \left(x^2+y^2\right)+x^2 y-\frac{y^3}{3} \nonumber\\
&+ \delta\left[x^4 y+x^2 y^3-y^5-\left(x^2+y^2\right)^2\right].
\end{align}

The introduction of the arbitrary parameter $\delta$ in (\ref{eq:2.5}) provides a valuable tool for studying the transition from the HH-system to the GHH-system, for this reason, all along this paper we shall consider values of the parameter $\delta$ in the interval $[0,1]$. 

\begin{table}
\caption{Critical values of energy  as a function of $\delta$ for bounded motion $E<E_{\rm min}$ and unbounded motion with three escape zones $E>E_{\rm max}$.}
\label{tab:1}       
\begin{tabular}{lll}
\hline\noalign{\smallskip}
$\delta$ & $E_{\rm min}$ & $E_{\rm max}$  \\
\noalign{\smallskip}\hline\noalign{\smallskip}
0 & 0.166666666666666667 & 0.166666666666666667 \\
0.1 & 0.0905432951155776544 & 0.101272469408731802 \\
0.5 & 0.0415033469448885181 & 0.0475015971035151890 \\
1 & 0.0267918006221522439 & 0.0304408354292176817 \\
\noalign{\smallskip}\hline
\end{tabular}
\end{table}

On the other hand, in order to classify the orbital motion as bounded or unbounded, we find two critical values of energy, $E_{\rm min}$ and $E_{\rm max}$, such that for energies below the threshold energy $E_{\rm min}$ there exists a closed zero-velocity surface, while for values larger than $E_{\rm max}$ the system shows three escape zones (see Table \ref{tab:1}). It should be clarified that $E_{\rm min}$ corresponds to the well-known energy of escape, so for values of $E$ larger than $E_{\rm min}$, there exists at least one escape channel.

\section{Bounded Orbits}
\label{sec3}

In the preceding section, it was possible to determine the value of the energy of escape for the GHH-system and hence to establish that the test particle could exhibit bounded motions for certain values of the total energy $E$, which depend on the values of the parameter $\delta$. In this section, we study the dynamical behavior of bounded orbits in four different cases $\delta = 0, 0.1, 0.5$ and 1, aiming to observe the incidence of the additional term in the HH-potential.

The system of equations (\ref{eqm1}-\ref{eqm4}), has been solved using a Runge-Kutta-Fehlberg algorithm RKF8(9) with variable time step. This method lets us numerically solve the system of equations once we know the constants $E$, $\delta$ and the initial conditions $x_0, y_0, p_{x_0},$  and $p_{y_0}$. The initial positions are chosen to fit the condition $U(x,y)=E$ for confined motion, specifically, we set $x_0=0.1$ and $y_0=0$ in all considered cases. To cover the whole phase space region of allowed motion, we used 15 different initial values of $p_{y}$ in the interval $[0, p_{max}]$, where $p_{max}$ corresponds to the largest value of $p_{y}$ that allows for a real numerical value of $p_{x}$. In each case, $p_{x}$ is determined by the integral of motion ${\cal H}=E$. The existence of the constant of motion indicates that the orbital motion takes place in a three-dimensional effective phase space in which the method of Poincar\'e sections is an adequate tool to characterize the motion between regular and chaotic.

\begin{figure*}[!t]
\centering
\resizebox{0.75\hsize}{!}{\includegraphics{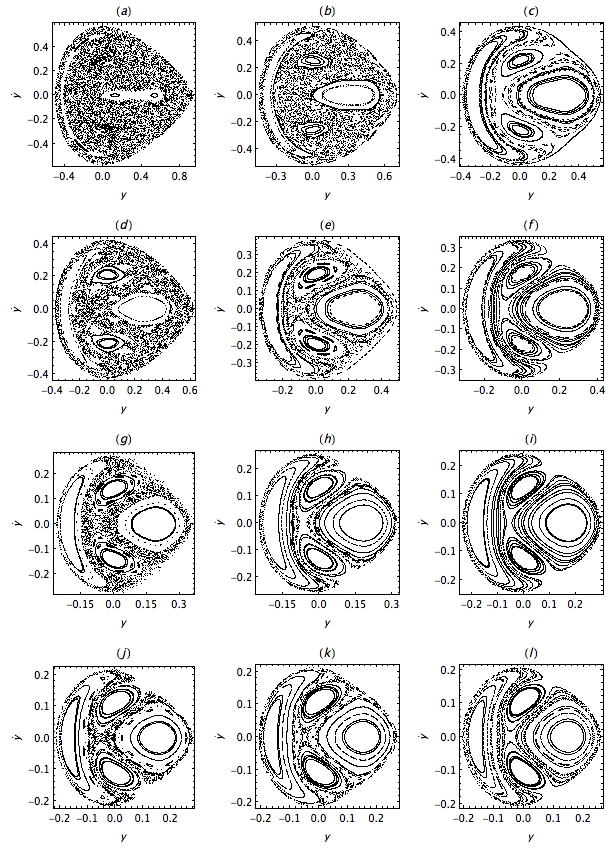}}
\caption{Poincar\'e surface of sections for  (a) $\delta=0$ and $n=1$, (b) $\delta=0$ and $n=21$, (c) $\delta=0$ and $n=41$, (d) $\delta=0.1$ and $n=1$, (e) $\delta=0.1$ and $n=16$, (f) $\delta=0.1$ and $n=31$, (g) $\delta=0.5$ and $n=1$, (h) $\delta=0.5$ and $n=11$, (i) $\delta=0.5$ and $n=21$, (j) $\delta=1$ and $n=1$, (k) $\delta=1$ and $n=6$, and (l) $\delta=1$ and $n=11$. In each panel, the system energy varies according to the relation $E=E_{\rm min}(1-n/100)$. The initial conditions have been set as $x_{0}=0.1$, $y_{0}=0$ varying $p_{y}$ in the interval $[0, p_{max}]$ and $p_{x}$ is determined by the energy conservation. Here, $p_{max}$ corresponds to the largest value of $p_{y}$ that allows for a real numerical value of $p_{x}$.}
\label{fig2}
\end{figure*}

In Fig. \ref{fig2}, we show the Poincar\'e sections for the GHH-system using different values of $\delta$ and gradually decreasing the value of $E$. It can be seen from panels $a$, $d$, $g$, and $j$ of Fig. \ref{fig2}  that values of $E$ very close to the critical energy $E_{\rm min}$ give rise to chaotic trajectories. On the other hand, for intermediate values of energy the system undergoes a transition from chaos to regularity (see panels $b$, $e$, $h$, and $k$), while in panels $c$, $f$, $i$, and $l$ of Fig. \ref{fig2} we show that no chaos seems to exist for values of $E$ much smaller than the critical energy  $E_{\rm min}$.  Additionally, it can be easily noted from the same figures that the chaos is weaker for larger values of $\delta$, in other words, the large chaotic zones in the H\'enon-Heiles case evolve into regular multi-island orbits endowed in a chaotic sea.

\begin{figure*}[!t]
\centering
\resizebox{0.75\hsize}{!}{\includegraphics{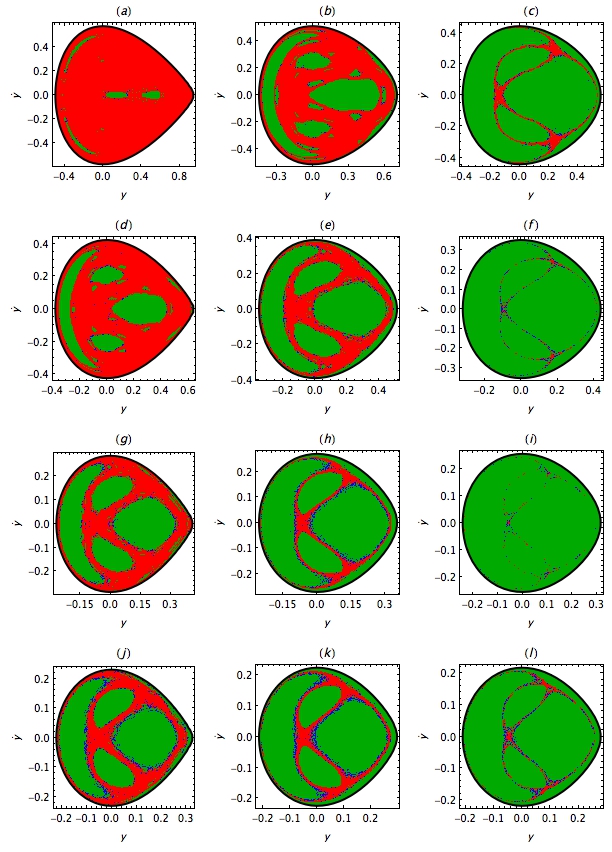}}
\caption{Orbital structure of the $(y,\dot{y})$ plane for (a) $\delta=0$ and $n=1$, (b) $\delta=0$ and $n=21$, (c) $\delta=0$ and $n=41$, (d) $\delta=0.1$ and $n=1$, (e) $\delta=0.1$ and $n=16$, (f) $\delta=0.1$ and $n=31$, (g) $\delta=0.5$ and $n=1$, (h) $\delta=0.5$ and $n=11$, (i) $\delta=0.5$ and $n=21$, (j) $\delta=1$ and $n=1$, (k) $\delta=1$ and $n=6$, and (l) $\delta=1$ and $n=11$. In each panel, the system energy varies according to the relation $E=E_{\rm min}(1-n/100)$. The color-code is as follows: regular orbits (green), sticky orbits (blue), chaotic orbits (red). (Color figure online)}
\label{fig21}
\end{figure*}

The classical Poincar\'e surfaces of section presented in Fig. \ref{fig2} give us a quick and rather general overview of the orbital structure in the GHH-system. However, if we require a more detailed view regarding the ordered or chaotic nature of the system then we have to perform a much more thorough scan of the available phase space by classifying large sets of initial conditions of orbits. In Fig. \ref{fig21} we use color-coded diagrams for revealing the orbital structure of the $(y,\dot{y})$ space. In particular, we define dense uniform grids of $300 \times 300$ initial conditions, regularly distributed inside the limiting curve, and we numerically integrate them for 5000 time units. For obtaining the character of the orbits we use the Smaller Alignment Index (SALI) \cite{Skokos2001}, which is defined as
\begin{equation}
\rm SALI(t) \equiv min(d_{-}, d_{+}),
\label{sali}
\end{equation}
where
\begin{align}
d_{-} & \equiv \left\Vert\frac{{\vec{w_1}}(t)}{\| {\vec{w_1}}(t) \|} - \frac{{\vec{w_2}}(t)}{\| {\vec{w_2}}(t) \|}\right\Vert, \nonumber\\
d_{+} & \equiv \left\Vert\frac{{\vec{w_1}}(t)}{\| {\vec{w_1}}(t) \|} + \frac{{\vec{w_2}}(t)}{\| {\vec{w_2}}(t) \|}\right\Vert,
\label{align}
\end{align}
are the alignments indices, while ${\vec{w_1}}(t)$ and ${\vec{w_2}}(t)$, are two deviation vectors which initially are orthonormal and point in two random directions. At every time step, each deviation vector is normalized to 1. Therefore, SALI is a dynamical quantity which inform us if the deviation vectors ${\vec{w_1}}(t)$ and ${\vec{w_2}}(t)$ have the tendency to obtain the same direction, either by coinciding or by becoming opposite. In the case of chaotic orbits the direction of the deviation vectors has the natural tendency to coincide with that of the most unstable nearby normally hyperbolic invariant manifold, which means that SALI tends to zero. On the contrary, in the case of regular orbits it lies on a torus which directly implies that the two deviation vectors ${\vec{w_1}}(t)$ and ${\vec{w_2}}(t)$ eventually become tangent to that torus, while generally they converge to entirely different directions. Consequently, the SALI fluctuates around a positive value.

Computationally, the nature of an initial condition is determined according to the final value of SALI at the end of the numerical integration. More precisely, if SALI $> 10^{-4}$ we have the case of a regular orbit, while if SALI $< 10^{-8}$ the orbit is chaotic. On the other hand, when $10^{-8} \leq \rm SALI \leq 10^{-4}$ we have the case of a sticky orbit\footnote{Sticky orbit refers to a special type of orbit which behaves as a regular one for a long-time interval and then it finally exhibits its true chaotic nature.}. From Fig. \ref{fig21}, it becomes evident that these color-coded diagrams clearly allow us to identify tiny local stability islands as well as weak chaotic layers, which are hard to be seen in a classical Poincar\'e surface of section.

\begin{figure}[!t]
\centering
\resizebox{.6\hsize}{!}{\includegraphics{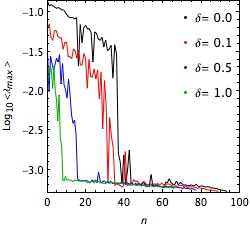}}
\caption{Energy dependence of $\lambda_{\rm max}$ in four different cases $\delta=0, 0.1, 0.5, 1$. In each case, the system energy varies in the interval $[0, E_{\rm min}]$ according to the relation $E=E_{\rm min}(1-n/100)$. (Color figure online)}
\label{fig3}
\end{figure}

A common criterion to measure the chaoticity of a dynamical system is to determine the maximum Lyapunov exponent $\lambda_{\rm max}$, which can be computed, for example, using the variational method \cite{Contopoulos1978} instead of the well-known two-particle approach, due to the fact that the last one could lead to inconsistent values of $\lambda_{\rm max}$ (see e.g. \cite{Dubeibe2014}). In Fig. \ref{fig3}, we validate the results obtained with the Poincar\'e sections by calculating the average $\lambda_{\rm max}$ of an ensemble of $10^5$ trajectories, in terms of the energy for different values of the parameter $\delta$. In this case, the largest Lyapunov exponent is calculated from the solution of the variational equations of the system, i.e., through the variational method. Because of the numerical nature of $\lambda_{\rm max}$, the Lyapunov exponents larger than the threshold value $10^{-3}$ are considered chaotic, while the ones below the threshold are considered regular.

Tracking the evolution of the system, by keeping the initial conditions fixed as $x_0 = 0.1, y_0 = -0.1$ and $p_{y_0} = 0.0$, it can be observed in Fig. \ref{fig3} that larger values of $\delta$ exhibit a smaller value of $\lambda_{\rm max}$, which also decay faster with decreasing energy (it should be noted that we have plotted $\lambda_{\rm max}$ against $n$, where the energy of the system varies according to the relation $E=E_{\rm min}(1-n/100)$). These results are in agreement with those obtained in Fig. \ref{fig2}, where the phase space is filled with regular islands for small energies.

\section{Unbounded Orbits}
\label{sec4}

\begin{figure*}[!t]
\centering
\resizebox{0.75\hsize}{!}{\includegraphics{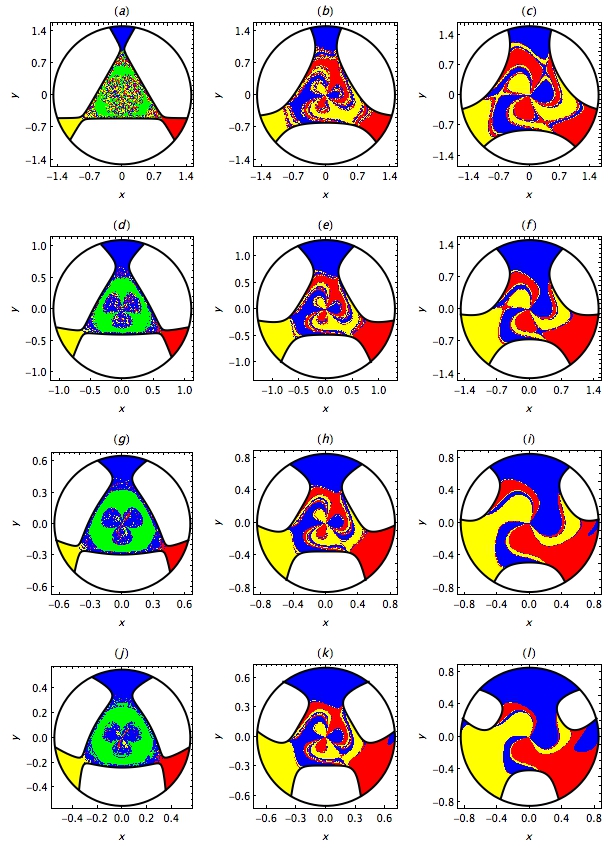}}
\caption{Exit basins of the GHH-Hamiltonian in the configuration space $(x,y)$ for: (a) $\delta=0$ and $n=2$, (b) $\delta=0$ and $n=50$, (c) $\delta=0$ and $n=200$, (d) $\delta=0.1$ and $n=2$, (e) $\delta=0.1$ and $n=50$, (f) $\delta=0.1$ and $n=200$, (g) $\delta=0.5$ and $n=2$, (h) $\delta=0.5$ and $n=50$, (i) $\delta=0.5$ and $n=200$, (j) $\delta=1$ and $n=2$, (k) $\delta=1$ and $n=50$, and (l) $\delta=1$ and $n=200$.  In each panel, the system energy varies according to the relation $E=E_{\rm max}(1+n/100)$. The color code is as follows: non-escaping orbits (green); escape through channel 1 (blue); escape through channel 2 (yellow); escape through channel 3 (red). The black circle denotes the scattering region. (Color figure online)}
\label{fig4}
\end{figure*}

\begin{figure*}[!t]
\centering
\resizebox{0.55\hsize}{!}{\includegraphics{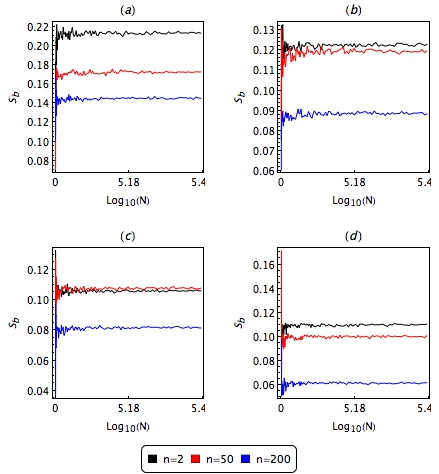}}
\caption{Basin Entropy $S_{b}$ as a function of the number of sub-matrices $N$ in four different cases: (a) $\delta=0$, (b) $\delta=0.1$, (c) $\delta=0.5$ and (d) $\delta=1$. In each panel, the system energy varies according to the relation $E=E_{\rm max}(1+n/100)$. (Color figure online)}
\label{fig5}
\end{figure*}

In this section, we study the trajectories in the GHH-system for energy values larger than the critical energy $E_{\rm max}$, in which the zero velocity surfaces are open with three escape channels. To do so, we solve numerically the system of differential equations (\ref{eqm1}-\ref{eqm4}) using a RKF-8(9) routine, with 90000 initial positions $(x_0, y_0)$ uniformly distributed along a square grid.  Following the usual convention for the escape basins of the HH-system, the initial momenta $(p_{x_0}, p_{y_0})$ are determined from  the conditions $\mathbf{r} \cdot \mathbf{p}=0$ and $\mathbf{r} \times \mathbf{p}>0$, with $\mathbf{r}=x \hat{\textbf{\i}}+y \hat{\textbf{\j}}$.

All trajectories are classified either into non-escaping or escaping orbits, where the last ones are subdivided according to the exit channel. As noted above, each trajectory can escape to infinity through three different exits, so we use the following convention: exit 1 $(y\rightarrow \infty)$, exit 2 $( x\rightarrow -\infty, y\rightarrow -\infty)$, and exit 3 $(x\rightarrow \infty, y\rightarrow -\infty)$. In Fig. \ref{fig4} we show the exit basin diagrams for the same values of the parameter $\delta$ used in section \ref{sec3}. In this plot, each initial condition is colored according to the escape channel through which it exits, i.e., escape through channel 1 with blue color, escape through channel 2 with yellow color and escape through channel 3 with red color. On the other hand, the green regions denote initial conditions of non-escaping orbits.

Following the same reasoning used in section \ref{sec3} to trace the evolution of the system, in Fig. \ref{fig4} we plot the exit basin for the GHH-system using different values of $\delta$ and gradually increasing the value of $E$. The first row (panels $a, b$, and $c$) corresponds to the HH-system $(\delta=0)$. In this case, the obtained results are in agreement with the literature \cite{deMoura1999,Zotos2017}, since the exit basins become smoother and well-defined as the energy increases. The same behavior is observed for $\delta=0.1$ (panels $d, e$, and $f$), $\delta=0.5$ (panels $g, h$, and $i$), and $\delta=1$ (panels $j, k$, and $l$). However, apparently the exit basins are smoother for larger values of $\delta$. In order to obtain conclusive results, the basin entropy for the invariant sets of the GHH-system will be thoroughly analyzed in the next paragraphs.

The basin entropy was introduced very recently \cite{Daza2016} as a new tool to quantitative measure the uncertainty of the basins. Here, the term uncertainty is understood as the difficulty to determine the final state to which a given initial condition will tend to. Unlike all the usual quantities used in nonlinear dynamics (Kolmogorov-Sinai entropy \cite{Kolmogorov1959,Sinai1959}, the topological entropy \cite{Adler1965}, or the expansion entropy \cite{Hunt2015}), the basin entropy refers to the topology of the basins instead of the evolution of the trajectories itself. For the sake of completeness, let us briefly describe the method for the calculation of the basin entropy.

Let us assume that our dynamical system has $N_{A}$ attractors (or final states) in a certain region $\Omega$ of the phase space, where $\Omega$ can be subdivided into a grid composed of $N$ square boxes with $\varepsilon^2$ trajectories per box.  Each box contains between 1 and $N_{A}$ final states, such that we can denote  $P_{i,j}$ as the probability that inside a box $i$ the resulting final state is $j$. Because the trajectories inside a box are independent, the Gibbs entropy of every box $i$ can be written as
\begin{equation}
S_{i}=\sum_{j=1}^{m_{i}}P_{i,j}\log\left(\frac{1}{P_{i,j}}\right)
\end{equation}
with $m_{i} \in [1,N_{A}]$ the number of final states inside the box $i$.

The entropy of the whole region $\Omega$  can be calculated as the sum of entropies of the resulting $N$ boxes of the grid, $S=\sum_{i=1}^{N} S_{i}$, thus the entropy relative to the total number of boxes $N$ (or basin entropy $S_{b}$) is given explicitly by the expression\footnote{For a detailed explanation of the method, we refer the interested reader to \cite{Daza2016}.}
\begin{equation}
S_{b}=\frac{1}{N}\sum_{i=1}^{N}\sum_{j=1}^{m_{i}}P_{i,j}\log\left(\frac{1}{P_{i,j}}\right).
\end{equation}

Since the final value for basin entropy $S_{b}$ strongly depends on the total number of boxes $N$ (such that for larger values of $N$ a more precise value of $S_{b}$ is obtained), here we use the approach presented in \cite{Daza2017} for the calculation of the basin entropy, in which $N$ squared-boxes are randomly selected in the phase space region $\Omega$ through  a Monte Carlo procedure. Following this method, in Fig \ref{fig5} we have computed the basin entropy $S_{b}$ as a function of the number of boxes $N$ for the exit basins presented in Fig \ref{fig4}. By choosing $\varepsilon=5$, it can be observed that in all cases the basin entropy tends to a constant value for $N> 10^{5}$. 

In panels $a, b, c$ and $d$ of Fig \ref{fig5} we compute the basin entropy for $\delta=0, 0.1, 0.5$ and 1, respectively. Additionally, in each panel the system energy varies as a function of $n$ according to the relation $E=E_{\rm max}=(1+n/100)$, i.e., we gradually increase the energy values starting very close to the respective critical energy $E_{\rm max}$. In analogy with the results of the previous section, it can be observed that as the energy gets closer to the critical energy $E_{\rm max}$ the basin entropy takes a larger value. Also, it is observed that larger values of $\delta$ significantly reduce the values of the basin entropy, which means that the uncertainty of the basins is larger for the HH-system than for the GHH-system.

\section{Conclusions}
\label{sec5}

In the present paper, we propose a new dynamical system that has as particular cases the H\'enon-Heiles and Verhulst Hamiltonians. As a key feature, our model has a finite energy of escape which allowed us to study the dynamics of bounded and unbounded orbits with three channels of escape. Via the Poincar\'e sections method and validated with the Largest Lyapunov exponent, we have shown that bounded orbits are mainly regular in all cases if the total orbital energy is much smaller than the escape energy. In the same vein, unbounded orbits were analyzed through the exit basin structure in configuration space and measuring the basin entropy. Here, we find that the exit basins become smoother and well-defined for larger values of the total orbital energy. Additionally, our numerical investigation suggests that the level of chaos as well as the uncertainty not only depend on the total orbital energy but also on the contribution of higher order terms in the series expansion, i.e, the orbital structure is much more regular in our model than in the one of H\'enon-Heiles, indicating that a third integral of motion seems to exist for energy values closer to the escape energy.

Finally, it deserves mentioning that the fractal structure of the basins was also analyzed by means of the boundary basin entropy $S_{bb}$. Our findings show that in all the considered cases the boundary basin entropy exhibits a similar tendency to the one observed for the basin entropy, such that the sufficient condition for the existence of fractal basin boundaries, $S_{bb}>\log(2)$, is always satisfied. This result is in agreement with previous studies indicating that these types of fractal structures appear not only in the paradigmatic H\'enon-Heiles system but also in a wide variety of open Hamiltonians with two degrees of freedom and three or more escape channels. We expect that our results will be useful in a wide variety complex systems studies, especially those related to the search of a third integral of motion in an axisymmetric potential.

\section*{Acknowledgements}
Our thanks go to the anonymous referee for their appropriate suggestions and for carefully reading the manuscript. One of the authors (FLD) gratefully acknowledges the financial support provided by Universidad de los Llanos and COLCIENCIAS, Colombia, under Grant No. 8840.

\end{document}